\documentstyle[12pt]{article}

\newcommand{\pslash}{P \hspace{-0.24cm} /}
\newcommand{\kslash}{k \hspace{-0.21cm} /}
\newcommand{\nslash}{n \hspace{-0.22cm} /}

\begin{document}

\title{\bf Lorentz invariance relations among parton distributions
 revisited}

\author{K.~Goeke$^a$, A.~Metz$^a$, P.V.~Pobylitsa$^{a,b}$ and M.V.~Polyakov$^{a,b}$
 \\[0.3cm]
$^a${\it Institut f\"ur Theoretische Physik II,} \\
{\it Ruhr-Universit\"at Bochum, D-44780 Bochum, Germany}\\[0.2cm]
$^b${\it Petersburg Nuclear Physics Institute, Gatchina,} \\
{\it St.\ Petersburg 188300, Russia}}

\date{\today}
\maketitle

\begin{abstract}
\noindent
We revisit the derivation of the so-called Lorentz invariance
relations between parton distributions.
In the most important cases these relations involve
twist-3 and transverse momentum dependent
parton distributions.
It is shown that these relations are violated if the path-ordered
exponential is taken into account in the quark correlator.
\end{abstract}

\noindent
{\bf 1.}~Parton distributions which are of higher twist and (or) dependent on
transverse parton momenta ($k_{\perp}$-dependent) contain important
information on the structure of the nucleon which is complementary to that
encoded in the usual twist-2 distributions.
Certain spin asymmetries in inclusive and semi-inclusive deep inelastic
scattering (DIS) as well as in the Drell-Yan process are governed by
twist-3 distributions~\cite{SLAC,levelt_94,jaffe_91}.
The $k_{\perp}$-dependent correlation functions typically give rise to
azimuthal asymmetries.
Very recently significant efforts have been devoted to measure such
asymmetries in semi-inclusive DIS~\cite{hermes_00,clas_03}.

In Refs.~\cite{tangerman_94,mulders_96,boer_98,boer_thesis} several
relations between twist-3 and (moments of) $k_\perp$-dependent parton
distributions have been proposed.
The derivation of these relations (called LI-relations in the
following) is based upon the general, Lorentz invariant decomposition
of the correlator of two quark fields, where the fields are located at
arbitrary space-time positions.
The LI-relations impose important constraints on the distribution
functions, which allow one to eliminate unknown structure functions in
favor of the known ones, whenever applicable.
Two specific LI-relations have been doubted in Ref.~\cite{kundu_01} by
an explicit calculation of the involved parton distributions in light
front Hamiltonian QCD using a dressed quark target. 
Although the arguments given in Ref.~\cite{kundu_01} are not complete,
that work motivated us to revisit the derivation of the LI-relations.
(Compare also the discussion in Ref.~\cite{ji_02b}.)

It is the purpose of the present Letter to study the validity of the
LI-relations in a model-independent way.
We find that they are violated if the proper path-ordered exponential
is taken into account in the quark correlation function.
The reason for this result lies in the fact that the gauge link
requires a decomposition of the correlator which contains more terms
than the ones given in Refs.~\cite{tangerman_94,mulders_96}.
Our result provides an explanation of the outcome of the 
model-calculation presented in Ref.~\cite{kundu_01}.

\noindent
{\bf 2.}~To begin with, we specify the correlation function through which the
$k_\perp$-dependent parton distributions are defined,
\begin{equation} \label{e:corr1}
\Phi_{ij}(x,\vec{k}_{\perp},S)  =
\int \frac{d \xi^- \, d^2 \vec{\xi}_{\perp}}{(2 \pi)^3} \,
 e^{i k^+ \xi^- - i \vec{k}_{\perp} \cdot \vec{\xi}_{\perp}} \,
 \langle P,S \, | \, \bar{\psi}_j(0) \,
 {\cal W}_1(0 , \xi) \,
 \psi_i(\xi) \, | \, P,S \rangle \biggr|_{\xi^+=0} \, .
\end{equation}
The target state is characterized by its four-momentum
$P = P^+ p + (M^2/2P^+) n$ and the covariant spin vector $S$
$(P^2 = M^2, \; S^2 = -1 , \; P \! \cdot \! S = 0)$, where the two
light-like vectors $p$ and $n$ satisfying
$p^2 = n^2 = 0$ and $p \! \cdot \! n =1$ have been
used.
The variable $x$ defines the plus-momentum of the quark via
$k^+ = x P^+$.
A contour leading to a proper definition of the $k_{\perp}$-dependent
parton distributions was given in
Refs.~\cite{collins_82,collins_02,ji_02,belitsky_02}\footnote{We note that the
choice of the contour depends on the process considered. Here we restrict
ourselves to the case of semi-inclusive DIS, although all our arguments 
are valid for other processes as well.}:
\begin{equation}
{\cal W}_1(0 , \xi)={\cal W}(0, \xi|n)\biggr|_{\xi^+=0}\, ,
\end{equation}
with
\begin{equation} \label{e:wilson1}
 {\cal W}(0, \xi|n) =
 [0,0,\vec{0} ; 0,\infty,\vec{0}]
 \mbox{} \times
 [0,\infty,\vec{0};\xi^+,\infty,\vec{\xi}_{\perp}]
 \mbox{} \times
 [\xi^+,\infty,\vec{\xi}_{\perp};\xi^+,\xi^-,\vec{\xi}_{\perp}] \,.
\end{equation}
In this equation, $[a^+,a^-,\vec{a}_\perp;b^+,b^-,\vec{b}_\perp]$ denotes
the Wilson line connecting the points $a^\mu=(a^+,a^-,\vec{a}_\perp)$
and $b^\mu=(b^+,b^-,\vec{b}_\perp)$ along a straight line.
It is important to note that the Wilson contour in eq.~(\ref{e:wilson1})  
not only depends on the coordinates of the initial and the final points 
but also on the light-cone direction 
$n$.\footnote{In fact Wilson lines that are near the 
light-cone rather than those exactly light-like are more appropriate
in connection with $k_{\perp}$-dependent parton
distributions~\cite{collins_82,collins_02}.
However, our general reasoning remains valid if we use a near light-cone
direction instead of $n$.}

The $k_{\perp}$-dependent parton distributions are defined
by the correlator in eq.~(\ref{e:corr1}) using suitable projections.
For instance, the unpolarized quark distribution is given by
$f_1(x,\vec{k}_{\perp}^2) = \rm{Tr}(\Phi \gamma^+)/2$.

Before dealing with the derivation of the LI-relations, we list the
most important examples~\cite{boer_98}, which will be shown to be not correct:
\begin{eqnarray} \label{e:rel1}
g_T(x) & = & g_1(x) + \frac{d}{dx} g_{1T}^{(1)}(x) \,,
 \\ \label{e:rel2}
h_L(x) & = & h_1(x) - \frac{d}{dx} h_{1L}^{\perp(1)}(x) \,,
 \\ \label{e:rel3}
f_T(x) & = & - \frac{d}{dx} f_{1T}^{\perp(1)}(x) \,,
 \\ \label{e:rel4}
h(x) & = & - \frac{d}{dx} h_1^{\perp(1)}(x) \,,
\end{eqnarray}
with
\begin{equation}
g_{1T}^{(1)}(x) = \int d^2 \vec{k}_{\perp} \,
 \frac{\vec{k}_{\perp}^2}{2 M^2} \, g_{1T}(x,\vec{k}_{\perp}^2) \,, \;
 etc.
\end{equation}
specifying the $k_{\perp}$-moments~\cite{mulders_96}.
All distributions on the {\it lhs} of eqs.~(\ref{e:rel1})--(\ref{e:rel4})
are of twist-3, while the functions on the {\it rhs} appear unsuppressed
in the observables.
For instance, $g_T$ is the well-known structure function measurable
via inclusive DIS on a transversely polarized target.
The functions $g_1$ and $h_1$, respectively, represent the quark
helicity and transversity distribution.
The distributions in eqs.~(\ref{e:rel1}),(\ref{e:rel2}) are time-reversal
even (T-even), while the ones in~(\ref{e:rel3}),(\ref{e:rel4}) are T-odd.
Only recently has it been explicitly shown that the
$k_{\perp}$-dependent T-odd distributions are non-vanishing in
general~\cite{collins_02,brodsky_02}.

\noindent
{\bf 3.}~The discussion of the LI-relations starts with the most general
correlator which, upon integration over $k^-$, reduces to the correlator
in eq.~(\ref{e:corr1}):
\begin{equation} \label{e:corr2}
\Phi_{ij}(P,k,S \, | n)  =
 \int \frac{d^4 \xi}{(2 \pi)^4} \, e^{i k \cdot \xi} \,
 \langle P,S \, | \, \bar{\psi}_j(0) \,
 {\cal W}(0,\xi| n) \, \psi_i(\xi) \, | \, P,S \rangle\, .
\end{equation}
We emphasize that the correlator~(\ref{e:corr2})
(like the one in~(\ref{e:corr1})) not only depends on the four-vectors
$P, \, k$ and $S$ but also, through the gauge link, on the light cone
direction $n$, which we have indicated now explicitly.
As we shall show in the following, it is precisely the presence of
this additional vector that spoils the LI-relations.

To write down the most general expression of the correlator in~(\ref{e:corr2}),
we impose the following constraints due to hermiticity and
parity\footnote{This implies a proper choice of the operator ordering in the
correlator (\ref{e:corr2}). The specific choice of this ordering is inessential
for our discussion. },
\begin{eqnarray}
\Phi^{\dagger}(P,k,S \, | n) & = &
 \gamma_0 \, \Phi(P,k,S \, | n) \, \gamma_0 \,,
\\
\Phi(P,k,S \, | n) & = &\gamma_0 \, \Phi(\bar P , -\bar S , \bar k \,| \bar n) \,
\gamma_0 \,,
\end{eqnarray}
where $\bar{P}^{\mu} = (P^0 , -\vec{P})$, etc.
With these constraints the most general form of the correlator reads
\begin{eqnarray} \label{e:decomp}
\Phi (P,k,S|n) &  = &
 M A_1 + \pslash \, A_2 + \kslash \, A_3
 + \frac{i}{2M} [\pslash,\kslash] A_4 + \ldots
\nonumber \\
 & & + \, \frac{M^2}{P \! \cdot \! n} \, \nslash \, B_1
 + \frac{i \, M}{2 \, P \! \cdot \! n} \, [\pslash,\nslash \, ] B_2
 + \frac{i \, M}{2 \, P \! \cdot \! n} \, [\kslash,\nslash \, ] B_3
 + \ldots \,,
\end{eqnarray}
where we have not listed those terms which only appear in the
case of target-polarization.
The structures in the second line in eq.~(\ref{e:decomp}) containing
the vector $n$ are absent in the decomposition given in
Refs.~\cite{tangerman_94,mulders_96}.
Note that, in order to specify the Wilson line in (\ref{e:corr2}), a rescaled
vector $\lambda n$ with some parameter $\lambda$ could be used instead of $n$.
By construction, the terms in~(\ref{e:decomp}) are not affected by such a
rescaling.

Next one makes use of the fact that integrating the correlator
(\ref{e:corr2}) upon $k^-$ necessarily leads to the correlator given in
(\ref{e:corr1}), i.e.,
\begin{equation} \label{e:identity}
\Phi (x,\vec{k}_{\perp},S) =
\int dk^- \Phi (P,k,S \, |n) \,.
\end{equation}
This identity has been used to derive the LI-relations.
As an explicit example, we consider the relation (\ref{e:rel4}) which
does not require target-polarization.
In this case, Eq.~(\ref{e:identity}) allows one to express the
involved distributions according to
\begin{eqnarray}
h_1^{\perp} (x,\vec{k}_{\perp}^2) & = & 2 P^+
 \int dk^- \, \bigg( - A_4 \bigg) \,,
 \\
h(x,\vec{k}_{\perp}^2) & = & 2 P^+
 \int dk^- \, \bigg( \frac{k \cdot P - x M^2}{M^2} \, A_4
         + (B_2 + x B_3) \bigg) \,.
\end{eqnarray}
If the structures in the second line in eq.~(\ref{e:decomp}) and,
hence, the amplitudes $B_i$ were absent then both $h_1^{\perp}$ and
$h$ would be given as an integral over the same amplitude $A_4$, which
is the origin of the LI-relation (see also in particular eq.~(2.30) 
in Ref.~\cite{tangerman_94} and Ref.~\cite{boer_thesis}).
However, as we have discussed, the amplitudes $B_2$ and $B_3$ need to
be taken into account as a direct consequence of gauge 
invariance.\footnote{Note that the amplitudes $B_i$ don't show up if
one connects the quark fields in the correlators in eqs.~(1),(9) by a
single straight Wilson line. However, one cannot define 
$k_{\perp}$-dependent parton distributions through such correlators.}
Accordingly, the relation~(\ref{e:rel4}) is violated.
One can easily extend our analysis to show that the
relations~(\ref{e:rel1})--(\ref{e:rel3}) involving target-polarization 
are violated also.

In summary we have shown that the so-called Lorentz invariance
relations between parton distributions are violated due to the
path-ordered exponential in the quark correlator.
We note that this result applies to the corresponding relations
among fragmentation functions as well.
\\[0.6cm]
\noindent
{\bf Acknowledgements:}
We are grateful to J.C.~Collins and A.V.~Efremov for discussions.
We thank V. Guzey for reading the manuscript.
The work of A.M. and M.V.P. has been supported by the Sofia Kovalevskaya
Programme of the Alexander von Humboldt Foundation.
The work has been partly supported by DFG, BMBF of Germany and the
project COSY-Juelich.


\end{document}